\documentclass[reprint,amsmath,amssymb,aps]{revtex4-1}
\usepackage{graphicx}% Include figure files
\usepackage{dcolumn}% Align table columns on decimal point
\usepackage{bm}% bold math
\usepackage{siunitx}
\usepackage{xcolor}
\usepackage{subfigure}
\def\build#1_#2^#3{\mathrel{\mathop{\kern 0pt#1}\limits_{#2}^{#3}}}

\begin{document}

\preprint{APS/123-QED}

\title{Lagrangian modeling of a non-homogeneous turbulent shear flow: Molding homogeneous and isotropic trajectories into a jet}
\author{Bianca Viggiano$^{\text{a,}}$}

 \email{bianca.viggiano@polymtl.ca}
\affiliation{
$^{\text{a}}$Department of Mechanical Engineering, Polytechnique Montr\'eal, Montr\'eal, QC H3C3A7, Canada}%
 \email{bianca.viggiano@polymtl.ca}

\author{Thomas Basset$^{\text{b}}$, Micka\"el Bourgoin$^{\text{b}}$, Ra\'ul Bayo\'an Cal$^{\text{c}}$, Laurent Chevillard$^{\text{b,d}}$, Charles Meneveau$^{\text{e}}$, Romain Volk$^{\text{b}}$}
\affiliation{
$^{\text{b}}$Univ Lyon, Ens de Lyon, Univ Claude Bernard, CNRS, Laboratoire de Physique, 46 all\'ee d’Italie F-69342 Lyon, France\\
$^{\text{c}}$Department of Mechanical Engineering, Portland State University, Portland, OR 97201, USA\\
$^{\text{d}}$Univ Lyon, Universit\'e Claude Bernard Lyon 1, CNRS UMR 5208, Institut Camille Jordan, 43 boulevard du 11 Novembre 1918, F-69622 Villeurbanne, France\\
$^{\text{e}}$Department of Mechanical Engineering, Johns Hopkins University, Baltimore, MD 21218, USA
}%

\date{\today}

\begin{abstract}
Turbulence is prevalent in nature and industry, from large-scale wave dynamics to small-scale combustion nozzle sprays. In addition to the multi-scale nonlinear complexity and both randomness and coherent structures in its dynamics, practical turbulence is often non-homogeneous and anisotropic, leading to great modeling challenges. In this letter, an efficient model is proposed to predict turbulent jet statistics with high accuracy. The model leverages detailed knowledge of readily available velocity signals from idealized homogeneous turbulence and transforms them into Lagrangian trajectories of a turbulent jet. The resulting spatio-temporal statistics are compared against experimental jet data showing remarkable agreement at all scales. In particular the intermittency phenomenon is accurately mapped by the model to this inhomogeneous situation, as observed by higher-order moments and velocity increment probability density functions. Crucial to the advancement of turbulence modeling, the transformation is simple to implement, with possible extensions to other inhomogeneous flows such as wind turbine wakes and canopy flows, to name a few.
\end{abstract}

\maketitle

%\tableofcontents

Turbulence continues to pose great conceptual and practical challenges in physics and engineering. Its most distinguishing characteristic is the turbulent kinetic energy cascade~\cite{richardson1922weather}, in which large eddies carry and transfer their energy onto a hierarchy of ever smaller eddies which are eventually dissipated by viscosity. This leads to enhanced diffusion~\cite{taylor1922diffusion} and even super-diffusion~\cite{richardson1926atmospheric} properties, crucial to mixing and dispersion. In a statistically homogeneous, isotropic and stationary turbulent (HIST) situation, fine scale properties of velocity fluctuations are well known. In particular, within the inertial range, the celebrated K41 self-similar theory~\cite{kolmogorov1941local} suggests that the moments of spatial velocity increments (\textit{structure functions}) over a given length-scale $\ell$ depend solely on the mean molecular energy dissipation rate $\langle \varepsilon\rangle$ and $\ell$, consistently with dimensional arguments. It was realized that this simple estimation fails when considering moments of order higher than three \cite{kolmogorov1962refinement, frisch1995turbulence,vassilicos2001intermittency}, and that this deviation is related to intermittency of the dissipation rate~\cite{kolmogorov1962refinement}.

In the Lagrangian frame, focusing on velocity temporal fluctuations along particles trajectories, the statistical study of turbulence is long standing~\cite{tennekes1972first,yeung2002lagrangian,toschi2009lagrangian,sawford2013lagrangian}, including generalizations of K41 dimensional arguments leading to multifractal models~\cite{byggstoyl1981closure, borgas1993multifractal,boffetta2002lagrangian,chevillard2003lagrangian,biferale2004multifractal,arneodo2008universal}. But this intermittent phenomenon affects our ability to accurately predict temporal fluctuations in chaotic systems~\cite{crisanti1993intermittency,duman2016dissipation} and thus, is a key factor to consider in risk assessment, and control and optimization schemes. In both the Eulerian (spatial) and Lagrangian (temporal) frameworks, the statistical knowledge developed in most of these studies focus on HIST flow. More recently, experimental and numerical investigations of Lagrangian trajectories in inhomogeneous situations have been proposed in the literature \cite{poulain2004spectral,ayyalasomayajula2006lagrangian,stelzenmuller2017lagrangian,shnapp2019extended,shnapp2021small,viggiano2021lagrangian}, thus calling on the development of models adapted to these more realistic situations than HIST.

The purpose of the present letter is to address this deficit in modeling and thus propose a quite general transformation which is able to map, or mold, Lagrangian velocity time series and their fine intermittent statistical structure, pertaining to HIST, towards a given inhomogeneous and anisotropic situation. A crucial step is the application of first and second-order statistics in the approximation of the inhomogeneous and anisotropic Eulerian advecting field, which are dependent on the position of particles in the flow, and can be modeled based on an input of a predefined HIST velocity signal. Therefore, the construction of our model differs from several previously proposed Lagrangian stochastic models, which were aimed at modeling realistic shear flows, which include spatial inhomogeneity \cite{legg1982markov,durbin1983stochastic,van1985random,wilson1981numericalp1,wilson1981numericalp2,wilson1981numericalp3,lipari2007numerical,minier2014guidelines} and are based on Gaussian noise and thus do not include intermittency (although some approaches with non-Gaussian forcing have also appeared~\cite{iliopoulos2004non}).  

The general premise of the proposed model is based on inverting a method presented by Batchelor \cite{batchelor1957diffusion} which was initially presented to facilitate analysis of inhomogeneous flows. Recall that a challenge for Lagrangian modeling in inhomogeneous flows is that, although the Eulerian flow field might be statistically stationary, Lagrangian trajectories are non-stationary. To overcoming this difficulty, Batchelor \cite{batchelor1957diffusion} presents a  stationarization technique to rescale the tracer velocity and time step along the trajectory to account for the evolution of Eulerian properties in the background of inhomogeneous flow fields. This method has been successfully applied to stationarize Lagrangian experimental jet data \cite{viggiano2021lagrangian}. Therefore, we propose to reverse the method in order to build up an inhomogeneous flow based on a stationary turbulence input. 
Here, the approach is illustrated while considering the case of a turbulent jet. It is based on a predefined signal and therefore can be easily implemented with a set of Lagrangian trajectories which are extracted, for instance, from DNS. The model, denoted herein as the \textit{Batchelor transformation}, is relatively easy to apply to any shear flow in which mean velocity and second-order statistics are known, thus covering many applications and situations of practical interest for natural and industrial flows.

In the Lagrangian framework, an ideal tracer path $\boldsymbol{X}(\tau)$, along its trajectory and starting at an initial position $\boldsymbol{X}(0)$, is related to the Eulerian velocity field $\boldsymbol{u}(\boldsymbol{x},t)$ of a given flow through the evolution,
\begin{equation}
      \frac{d\boldsymbol{X}(\tau)}{d\tau} = \boldsymbol{u}[\boldsymbol{X}(\tau),\tau] \equiv \boldsymbol{v}({\tau}),\label{eq:path}
\end{equation}
\noindent where we have introduced the Lagrangian velocity vector $\boldsymbol{v}(\tau)$ at some time $\tau$. Henceforth, we will denote the HIST Lagrangian velocity vector as $\tilde{\boldsymbol{v}}(\tilde{\tau})$, located at the position $\tilde{\boldsymbol{X}}(\tilde{\tau})$, which is obtained while considering a HIST advecting Eulerian field $\tilde{\boldsymbol{u}}(\boldsymbol{x},t)$ in Eq. \ref{eq:path}. Therefore, we keep the notation $\boldsymbol{v}(\tau)$ for Lagrangian velocities in a (modeled) round jet. 
In the sequel and without loss of generality, we consider nondimensionalized HIST trajectories such that $\langle \tilde{\boldsymbol{v}}(\tilde{\tau}) \otimes\tilde{\boldsymbol{v}}(\tilde{\tau})\rangle=\boldsymbol{I}$, where $\langle \bullet\rangle$, $\otimes$ and $\boldsymbol{I}$ stand for respectively the empirical average over the ensemble of trajectories, the tensor product and the identity matrix. The nondimensional time step, $\tilde{\tau}$, is expressed in units of the integral time scale. Notice that for HIST, components of $\tilde{\boldsymbol{v}}$ are independent up to second-order and in this nondimensionalized framework, are of unit-variance.

Since it is tremendously difficult to measure and/or simulate the spatio-temporal advecting velocity field $\boldsymbol{u}(\boldsymbol{x},t)$ of a round jet entering in Eq. \ref{eq:path}, at any time and any position, the so-called  \textit{Batchelor transformation} consists, instead, of solving Eq. \ref{eq:path}, in mapping a set of HIST velocities $\tilde{\boldsymbol{v}}(\tilde{\tau})$ (easily obtained from, for instance, open-access DNS), through a non-linear transformation. To build up this transformation for application to the specific flow of a turbulent jet in which we want to preserve symmetry and introduce cross-correlation, it will be convenient to express the Cartesian fluid particle coordinates $\boldsymbol{X}(\tau)= \left(X_x(\tau),X_y(\tau),X_z(\tau)\right)$ in the cylindrical system with the streamwise direction of the jet as the principal axis such that $\boldsymbol{X}(\tau)=  \left(r(\tau),\theta(\tau),z(\tau)\right)$, with $r(\tau)= \sqrt{X_x^2(\tau)+X_y^2(\tau)}$, $\theta(\tau) =\text{atan2}(X_y(\tau),X_x(\tau))$ and $z(\tau) =  X_z(\tau)$, where  $\text{atan2}$ is a common variation on the arctangent function.

The full transformation of the given ensemble of HIST velocities $\tilde{\boldsymbol{v}}(\tilde{\tau})$ into the Lagrangian velocity of a jet, $ \boldsymbol{v}(\tau)$, reads in Cartesian coordinates as
\begin{equation}\label{eq:TransformHISTtoJet}
 \boldsymbol{v}(\tau)\build{=}_{}^{\text{model}} \boldsymbol{\mathcal R}^{-1}_{\theta(\tau)}\left(\boldsymbol{\mathcal L}_{ \boldsymbol{X}(\tau)}\boldsymbol{\mathcal R}_{\theta(\tau)}\tilde{\boldsymbol{v}}(\tilde{\tau})+ \boldsymbol{\overline{u}}_{\boldsymbol{X}(\tau)} \right),
\end{equation}
\noindent where the structure of the jet is introduced through the deterministic vector $\boldsymbol{\overline{u}}_{\boldsymbol{X}(\tau)}$ and matrix $\boldsymbol{\mathcal L}_{ \boldsymbol{X}(\tau)}$. 
The rotation matrix $\boldsymbol{\mathcal R}_{\theta(\tau)}$ of the angle $\theta(\tau)$ along the axis of the jet is also introduced and allows to easily jump between Cartesian and cylindrical frames. Making the crude approximation that for a given set of random instances of HIST velocities $\tilde{\boldsymbol{v}}(\tilde{\tau})$, the position $\boldsymbol{X}(\tau)$ of particles in the jet are statistically independent  of $\tilde{\boldsymbol{v}}(\tilde{\tau})$, the deterministic parameter functions $\boldsymbol{\overline{u}}_{\boldsymbol{X}(\tau)}$ and $\boldsymbol{\mathcal L}_{ \boldsymbol{X}(\tau)}\boldsymbol{\mathcal L}_{ \boldsymbol{X}(\tau)}^\top$ ($^\top$ stands for matrix transpose) coincide respectively with the Eulerian average and covariance matrix of the vector velocity field in the cylindrical frame at the position $\boldsymbol{X}(\tau)$. Taking into account the full nonlinear underlying evolution obtained while solving the dynamics given in Eq. \ref{eq:path} with the modeled Lagrangian velocity $ \boldsymbol{v}(\tau)$ (Eq. \ref{eq:TransformHISTtoJet}), and therefore making  $\boldsymbol{X}(\tau)$ a functional of the set  $\tilde{\boldsymbol{v}}(\tilde{\tau})$, is for the moment out of reach from a theoretical side.

For the proposed application to a round turbulent jet, these required deterministic quantities are well known from previous studies~\cite{pope2000turbulent}. In Eq.~\ref{eq:TransformHISTtoJet}, at a given position expressed in the cylindrical frame $\boldsymbol{X}=(r,\theta,z)$, the Eulerian mean velocity in cylindrical coordinates in a turbulent jet without swirl is 
$\boldsymbol{\overline{u}}_{\boldsymbol{X}}= [\overline{u}_{r}\left(r,z\right), 0,   \overline{u}_{z}\left(r,z\right)]^T$
and the matrix $\boldsymbol{\mathcal L}_{ \boldsymbol{X}(\tau)}$, obtained as a Cholesky decomposition of the Eulerian covariance matrix, can be written as
\begin{align}\label{eq:DefLX}
 \boldsymbol{\mathcal L}_{ \boldsymbol{X}}=\begin{bmatrix} \sigma_{r}\left(r,z\right)& 0& 0\\
   0 & \sigma_{\theta}\left(r,z\right) &0\\ 
    \frac{f_{rz}\left(r,z\right)}{\sigma_{r}\left(r,z\right)}& 0& \sqrt{\sigma_{z}^2\left(r,z\right)-\frac{f_{rz}^2\left(r,z\right)}{\sigma^2_{r}\left(r,z\right)}}
    \end{bmatrix},
\end{align}
where the $\tau$ dependence is dropped to be concise. The remaining average Eulerian velocity profile functions $\overline{u}_{z}$ and $\overline{u}_{r}$ as well as the radial $\sigma_r$, azimuthal $\sigma_\theta$ and axial $\sigma_z$ standard deviations, and the radial/axial $f_{rz}$ cross-correlation, are provided in Supplemental Materials. Because of symmetry constraints, they are all functions of only the centerline velocity, $U_{0}(z)$, and the self-similarity variable $\eta = r/(z-z_0)$, where $z_0$ is the virtual origin. In a round jet, they are all independent of $\theta$ for statistical symmetry reasons, but it nonetheless enters the formulation as soon as Lagrangian velocities are expressed in the Cartesian frame.

As the last step entering in the transformation, the expression of the local time $\tau(\tilde{\tau})$ in the jet as a function of the nondimensional time $\tilde{\tau}$ of the set of HIST velocities has to be stated. It is obtained by integrating the differential relation
$d\tau = T_{E_z}\left[X_z(\tau)\right]d\tilde{\tau}$, with the initial condition $\tau(\tilde{\tau}=0)=0$. Here, the dimensional large eddy turnover timescale, $T_{E_z}$, which is assumed to depend solely on the axial coordinate is also needed, as part of the \textit{Batchelor transformation} method (see Supplemental Materials). Positions, $\boldsymbol{X}\left( \tau\left(\tilde{\tau}\right)\right)$, are numerically estimated at each local time step $\tau(\tilde{\tau})$.

For the presentation of the model, velocities $\tilde{\boldsymbol{v}}(\tilde{\tau})$ along 32$^3$  HIST trajectories are extracted from DNS of forced isotropic turbulence provided by the Johns Hopkins Turbulence Database (see http://turbulence.pha.jhu.edu). The dataset has a Reynolds number based on the Taylor microscale of $\mathcal R_\lambda$~=~418. It is on a 1024$^3$ periodic grid and data is stored at a given constant resolution $\Delta \tilde{t} = 0.002$. Details on how to generate the Lagrangian trajectories from the Eulerian field are found in \citet{li2008public} and \citet{yu2012studying}. In addition, to validate the model predictions with data, we use experimental Lagrangian jet data~\cite{viggiano2021lagrangian}. More information on the experimental procedures is provided in the Supplemental Materials text.

\begin{figure}
    \centering
   \includegraphics[width=8.6cm]{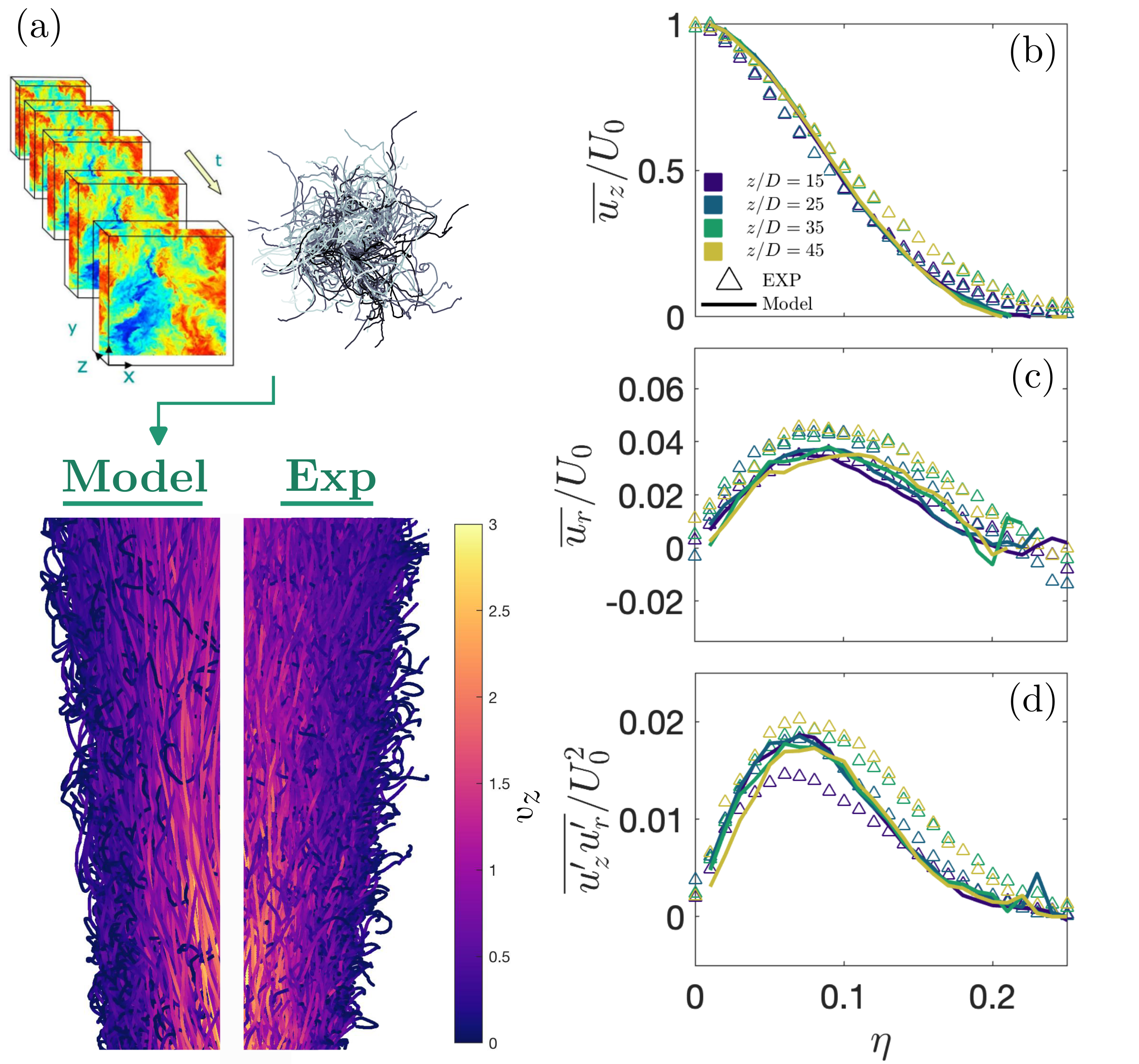}
    \caption{(a) The proposed modeling technique; obtain HIST trajectories from an online database, such as JHTDB, to input into the \textit{Batchelor transformation} resulting in Lagrangian trajectories which are compared with experimental trajectories. Self similar profiles of the mean (b) axial and (c) radial velocity as well as (d) the shear stress.}
\label{fig:jetschematic}
\end{figure}

An overview of the methodology is presented in Fig. ~\ref{fig:jetschematic}(a).
We include the initial input (Lagrangian time histories of turbulence along fluid trajectories in homogeneous isotropic turbulence obtained from fluid particle tracking in Eulerian DNS). These HIST time series of velocity along Lagrangian paths are rescaled through the \textit{Batchelor transformation} to comply with known statistical features of a turbulent round jet. Once rescaled, they are used to integrate particle positions and can be used to generate many realizations of such trajectories which can then be compared to Lagrangian data from experiments (bottom of the Fig.~\ref{fig:jetschematic}(a)). Qualitatively, this provides clear visual similarities between the Lagrangian jet trajectories experimentally measured and those which have been reconstructed by instationarizing a HIST time series from the DNS. The colorbar denotes the magnitude of the axial velocity, which is consistent between the modeled and experimental jet in its axial and radial decay. Furthermore, the shapes of the two jets are very similar. 

The figure also presents basic large scale inhomogeneous Eulerian statistics;~\ref{fig:jetschematic}(b) mean axial velocity,~\ref{fig:jetschematic}(c) mean radial and~\ref{fig:jetschematic}(d) mean shear stress profiles. These profiles verify that the spatial attributes of the reconstructed jet are quantitatively behaving as expected. To calculate spatial averages of the Lagrangian data, the modeled and experimental data are binned axial and radially, specifically in $z$ and $\eta=r/z$ in increments of $\SI{1}{mm}$ and 0.01, respectively. The time and ensemble average is then taken with respect to the quantity of interest.

Each of the modeled profiles, presented as solid lines, follow trends nicely with the experimental data (symbols) for each the considered downstream locations, $z/D=$ 15, 25, 35 and 45. The mean axial and radial velocity profiles, Figs.~\ref{fig:jetschematic}(b) and~\ref{fig:jetschematic}(c), display a nice collapse for the given downstream locations of interest. The similarity observed in the radial velocity profiles provides further validation of the model due to the fact that the experiment is only seeded from a point source (the nozzle), the model is also fed from the nozzle location and therefore the ambient fluid is not `tagged' and the retrieved radial profiles correctly reflect the nozzle seeding~\cite{basset2022entrainment}. The shear stress, Fig.~\ref{fig:jetschematic}(d), is also provided to show that the model based on the diagonalization of the covariance provides realistic predictions, as discussed in Supplemental Material. Here again, the model reproduces our experimental findings with success. Other large scale Eulerian quantities (not shown) such as standard deviations and concentration profiles of the seeding reveal similar trends. The minimal differences observed can be attributed to the trajectory length which is limited by the simulation time, which does not allow the full development of the jet to largest $\eta$ values. 

The novelty of this approach is herein demonstrated by its ability to accurately model the full multi-scale dynamics associated with Lagrangian statistics including the small scale intermittency. This model feature is achieved based on the non-Gaussianity of small scale velocity increments, particularly the skewness of Lagrangian increments (which is specific to inhomogeneous turbulence and not present in the original HIST signal) and their flatness. It is noted that due to the inhomogeneity of the jet (both experimental and modeled) statistics are estimated from trajectories which are conditioned based on their initial location. Further detail on the sampling is provided in Supplemental Material. 

Fig.~\ref{fig:corr_s2_s4}(a) and~\ref{fig:corr_s2_s4}(b) present the correlation of axial and radial velocity, respectively, where, for example, the axial velocity correlation, $\mathcal C_{v_{z}}(\tau) = \langle v_z(t_0+\tau)v_z(t_0)\rangle$. The correlations are given as a function of the normalized time delay $\tau U_J/D$ (where $U_J$ is the nozzle jet speed and $D$ is the nozzle diameter). It is noted here that the time-scale is adjusted at the first location ($z/D = 15$) to allow a temporal collapse of the data based on the correlation of velocity. This temporal discrepancy could be due to the fact that self-similarity is not yet reached at the nearest to the nozzle location. For both velocities, the dynamics are well described by the model at small and large time separations. Small deviations are present at the largest $\tau$ values due to the low convergence of the experimental data as the length of the track increases.

\begin{figure}
    \centering
   \includegraphics[width=8.6cm]{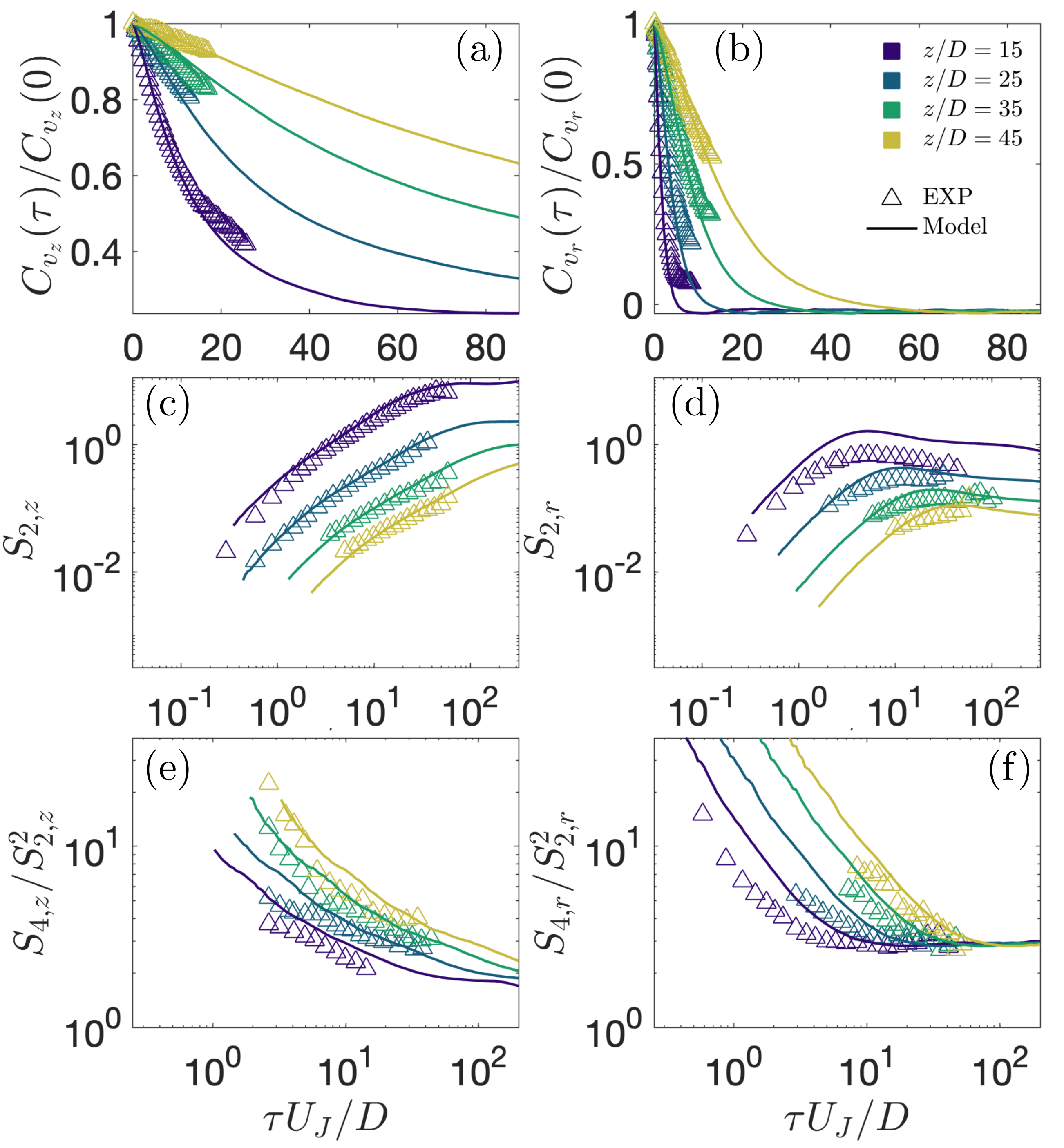}
    \caption{Lagrangian two-time statistics of the model and experiment at 4 locations downstream: (a) axial and (b) radial velocity correlations, (c) axial and (d) radial second-order structure function and the flatness for the (e) axial and (f) radial components. The input is trajectories obtained from the DNS.}
\label{fig:corr_s2_s4}
\end{figure}

Next, structure functions are used to validate the model output against the experimental measurements and are calculated as $S_{n,i}(\tau) = \langle [v_i(t_0+\tau)-v_i(t_0)]^n\rangle$, where $i$ is the velocity component and $n$ represents the moment order. The second-order structure function is provided in Figs.~\ref{fig:corr_s2_s4}(c) and~\ref{fig:corr_s2_s4}(d), shown in log units to visualize small scale phenomena. Generally an agreement is observed for the axial and radial velocity with some differences noted at $z=15D$. This is possibly due to the fact that self-similarity might not be fully developed at that location for the experimental data, as previously noted. Otherwise, the curves collapse well within the range obtainable from the experimental data. The model extends the accessible ranges of the scales within the jet, providing the well known -2 power law at the dissipate scales~\cite{yeung1989lagrangian}.

The most notable method to quantify the intermittency effects of a given signal is through higher-order moment analysis such as the 4th-order structure function, presented in Figs.~\ref{fig:corr_s2_s4}(e) and~\ref{fig:corr_s2_s4}(f) as flatness, $S_{4,i}/S_{2,i}^2$. A purely Gaussian signal would present a constant value of three. At large scales, this Gaussian response is observed, but as $\tau\rightarrow 0$, intermittency becomes crucial to the signal dynamics and there exists a steep incline to a plateau in the far dissipative range. The 4th-order moment is not easily obtained within experiments and therefore a very limited range of experimental data is presented. The model agrees well with the limited experimental data profiles, especially for $z>25D$ for both $v_z$ and $v_r$ and again, the model is able to capture a wider range of scales than those acquired by the experiments.

\begin{figure}[!htb]
    \centering
   \includegraphics[width=8.6cm]{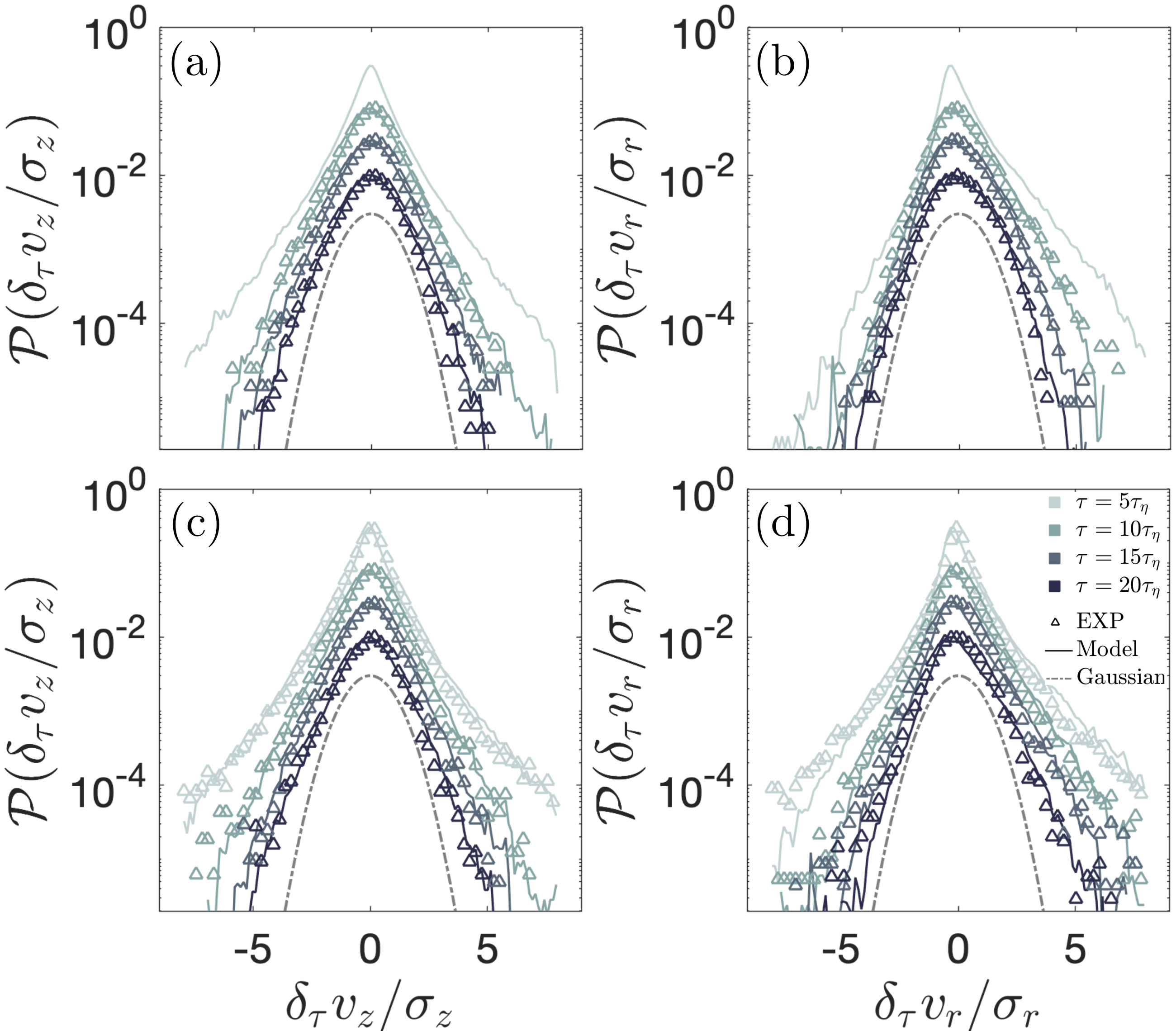}
    \caption{PDF of the velocity increment of the model and the experiment for various $\tau$ separations. PDFs of the (a) axial and (b) radial velocity increments at $x/D$ = 15 and (c) axial and (d) radial velocity PDFs at $x/D$ = 45.}
\label{fig:PDF}
\end{figure}

An alternative method to study the potential for intermittency to be reproduced by the model is by investigating the small scale dynamics via PDFs of the Lagrangian velocity increments. By considering separately the PDFs for different time separation, $\tau$, we can isolate the intermittent behaviors and see how well the model captures these unique attributes of turbulence for a given scale. Fig.~\ref{fig:PDF} presents the PDF of the velocity increment for the axial, $\delta_\tau v_z$, and radial, $\delta_\tau v_r$, components, normalized by their respective standard deviation. Multiple scale separations are included at the two downstream locations presented, denoted by the bounding box. The scale separations range from very small, $\tau = 5\tau_\eta$, to nearly the integral timescale at $\tau = 20\tau_\eta$. The experimental data was unobtainable at the smallest probed time scales at $z/D=15$ and is omitted. Finally, a Gaussian PDF is also included in each figure for reference and the curves are arbitrarily shifted for clarity.

The axial velocity PDFs at both downstream locations show similar trends. The largest scale curves ($\tau=20\tau_\eta$) nearly reach the Gaussian distribution in shape and as $\tau$ decreases, the tails of the PDFs become more prominent. The model is in almost absolute agreement with the experiment, with small exceptions at the edges of the PDFs when the intermittency is not perfectly captured, as was also observed in the flatness representations. The radial velocity increments present a positive skewness in the shape of the PDFs for all locations and $\tau$ separations, which appears most drastic nearest the jet exit. 

The modeling of the jet captures the trend of the smaller scales presenting over-Gaussian behavior and the larger scales exhibiting near Gaussian dynamics. It furthermore captures this very distinct behavior of the positive skewness with curves collapsing nicely for nearly all presented data. This indicates that the model can replicate the subtle details of the jet and the intermittency contributions over a large scale range, as the jet develops.

In this letter, a model is proposed to build up a set of Lagrangian trajectories from a homogeneous, isotropic and stationary signal. To summarize, results show that it is possible to cast an anisotropic, inhomogeneous (and hence non-stationary from the Lagrangian point of view) turbulent flow as complex as a jet from a generic input of homogeneous and isotropic turbulence, using a simple ``mold'' which only features low-order and large scale properties of the jet, namely the mean and variance Eulerian velocity profiles. In doing so, a very realistic fine scale statistical structure of the jet molded trajectories is found, including appropriate intermittent features. Future investigations will be devoted to study the implications of such a transformation (Eq. \ref{eq:TransformHISTtoJet}) on stochastic models of HIST which may include finite Reynolds number and intermittency effects~\cite{sawford1991reynolds,viggiano2020modelling} and those recently obtained from machine learning techniques \cite{li2023synthetic}.  The ability to model a jet has far-reaching implications with its application to numerous industrial and natural flows such as volcanic eruptions, pathogen spread via coughing, and nozzle sprays. Such an approach opens an important new perspective to the accurate modeling of inhomogeneous (realistic) turbulence and leads to the question: Can this ability to capture the multi-scale complex features of turbulence in the jet be applied to a broader class of inhomogeneous flows? Extensions of this technique to other inhomogeneous flows, both self-similar and not, leaves many open problems to be investigated. 

\begin{acknowledgments}
B.V. and R.B.C. acknowledge the National Science Foundation (grant GEO-1756259). R.B.C. is also thankful for the support provided through the Fulbright Scholar Program and NSF-PMP-2223235. B.V., T.B., R.V. and M.B. acknowledge support from the Initiative d’excellence de Lyon (IDEXLYON) of the University of Lyon in the framework of the Programme Investissements d’Avenir (ANR-16- IDEX-0005). L.C. is partially funded by Agence National de la Recherche ANR-20-CE30-0035 and by the Simons Foundation Award ID 651475.  C.M. acknowledges support from the National Science Foundation, which also supports the JHTDB infrastructure (grant CSSI-210387).
\end{acknowledgments}

\newpage
\begin{widetext}
\subsection*{Lagrangian modeling of a non-homogeneous turbulent shear flow: Molding homogeneous and isotropic trajectories into a jet - Supplemental Material}%

\section*{Implementation of the model}
The model execution is performed on-the-fly, meaning that at each time step, an iterative process is implemented to obtain the position of the tracer particle at a given time step $\tau$. Recall first that the local time of the jet is a function of the non-dimensionalized input time $\tilde{\tau}$. In practice this is estimated by a discretization of $d\tau = T_{E_z}\left[X_z(\tau)\right]d\tilde{\tau}$ where the integral time $T_{E_z}$ is a function of only the axial position in the jet in the given formulation (relation is presented below). Note that $\tau(\tilde{\tau})$ is not a linear function due to the evolving background temporal Eulerian field $T_{E_z}$. Note also that the inputted time $\tilde{\tau}$ is a non-dimensional time either coming directly from the simulation or normalized, for example by the large eddy turnover time.

This iterative calculation of position of each fluid element is obtained by numerically estimating
\begin{equation}\label{eq:IntegralOverTildeS}
	\boldsymbol{X}\left( \tau\left(\tilde{\tau}\right)\right)-\boldsymbol{X}\left( 0\right)=\int_{\tilde{s}=0}^{\tilde{\tau}}
	\boldsymbol{v}\left( \tau\left(\tilde{s}\right)\right)T_{E_z}\left[X_z\left( \tau\left(\tilde{s}\right)\right)\right]d\tilde{s},
\end{equation}

\noindent recalling here that the modeled Lagrangian velocity $ \boldsymbol{v}\left( \tau\right)$ is provided by 

\begin{equation}\label{eq:TransformHISTtoJet2}
	\boldsymbol{v}(\tau)\build{=}_{}^{\text{model}} \boldsymbol{\mathcal R}^{-1}_{\theta(\tau)}\left(\boldsymbol{\mathcal L}_{ \boldsymbol{X}(\tau)}\boldsymbol{\mathcal R}_{\theta(\tau)}\tilde{\boldsymbol{v}}(\tilde{\tau})+ \boldsymbol{\overline{u}}_{\boldsymbol{X}(\tau)} \right),
\end{equation}

\noindent where the matrix $\boldsymbol{\mathcal L}_{ \boldsymbol{X}(\tau)}$ and vector $\boldsymbol{\overline{u}}_{\boldsymbol{X}(\tau)}$ are prescribed based on known relations (defined below). Finally we note that this equation depends explicitly on the position $\boldsymbol{X}\left( \tau\right)$ of the fluid element tracers.

\section*{Notes on the model inputs and their nonlinearity}

The nonlinear nature of the \textit{Batchelor transformation} arises from the fact that the positions $\boldsymbol{X}(\tau)$ are determined by  ${d\boldsymbol{X}(\tau)}/{d\tau} = \boldsymbol{u}[\boldsymbol{X}(\tau),\tau] \equiv \boldsymbol{v}({\tau})$ and are therefore functionals of the set of input unit variance velocities, $\tilde{\boldsymbol{v}}$, in an intricate way. Notice that the projection in the cylindrical frame of the set of HIST velocities $\tilde{\boldsymbol{v}}$ preserves a simple covariance structure of the angle, $\theta(\tau)$, and is assumed to be distributed independent of it. Specifically, $\langle  \boldsymbol{\mathcal R}_{\theta(\tau)}\tilde{\boldsymbol{v}}(\tilde{\tau}) \otimes\boldsymbol{\mathcal R}_{\theta(\tau)}\tilde{\boldsymbol{v}}(\tilde{\tau})\rangle=\langle  \boldsymbol{\mathcal R}_{\theta(\tau)}\left( \tilde{\boldsymbol{v}}(\tilde{\tau}) \otimes\tilde{\boldsymbol{v}}(\tilde{\tau})\right)\boldsymbol{\mathcal R}_{\theta(\tau)}^\top\rangle\approx \langle  \boldsymbol{\mathcal R}_{\theta(\tau)}\boldsymbol{I}   \boldsymbol{\mathcal R}_{\theta(\tau)}^\top\rangle=\boldsymbol{I}$, where $\otimes$ is the matrix product and $\boldsymbol{I}$ the identity matrix. In reality the situation is more complicated (the angle $\theta(\tau)$ is a nonlinear function of the history of HIST velocities $\tilde{\boldsymbol{v}}(\tilde{\tau})$). For this reason, a more precise estimation of the covariance of the projected velocities is difficult to obtain and out of the scope of this project. Nonetheless, because $\theta(\tau)$ is a random variable,  $\boldsymbol{v}(\tau)$ is not expected to be a Gaussian process, even if $\tilde{\boldsymbol{v}}(\tilde{\tau})$ is assumed to be one.

\section*{Model input parameters}

For completeness, we provide the details of the various elements of the vector $\boldsymbol{\overline{u}}_{\boldsymbol{X}(\tau)}$ and the matrix  $\boldsymbol{\mathcal L}_{ \boldsymbol{X}(\tau)}$ used as inputs into the model based on the location of the particle in $\eta$, i.e., in $r$ and $z$. Recall here the components of these inputs are:

\begin{equation}\label{eq:DefmX}
	\boldsymbol{\overline{u}}_{\boldsymbol{X}}= \begin{bmatrix} 
		\overline{u}_{r}\left(r(\tau),z(\tau)\right)\\
		0\\
		\overline{u}_{z}\left(r(\tau),z(\tau)\right)
	\end{bmatrix},
\end{equation}

\noindent and
\begin{equation}\label{eq:DefLX}
	\boldsymbol{\mathcal L}_{ \boldsymbol{X}(\tau)}=\begin{bmatrix} \sigma_{r}\left(r(\tau),z(\tau)\right)& 0& 0\\
		0 & \sigma_{\theta}\left(r(\tau),z(\tau)\right) &0\\ 
		\frac{f_{rz}\left(r(\tau),z(\tau)\right)}{\sigma_{r}\left(r(\tau),z(\tau)\right)}& 0& \sqrt{\sigma_{z}^2\left(r(\tau),z(\tau)\right)-\frac{f_{rz}^2\left(r(\tau),z(\tau)\right)}{\sigma^2_{r}\left(r(\tau),z(\tau)\right)}}
	\end{bmatrix}.
\end{equation}

First, the self-similar nature of the jet~\cite{hussein1994velocity,so1986similarity}, allows the time-averaged streamwise velocity profile to be written as,

\begin{equation}\label{eq:DefMeanAxialComp}
	\frac{\overline{u}_z(r,z)}{U_0(z)} = f(\eta).
\end{equation}

\noindent Here the centerline velocity is introduced $U_0(z)= \overline{u}_z(r=0,z)$ as well as the self-similarity variable, $\eta$, where $\eta = r/(z-z_0)$ and $z_0$ is the virtual origin of the jet. The centerline mean velocity is known to follow an inverse power-law with distance according to~\cite{hussein1994velocity}

\begin{equation}
	U_{0}(z)={BU_JD}/({z-z_0})
\end{equation}
\noindent where $U_J$ the exit velocity at the nozzle, $D$ is the nozzle diameter and $B$ is the dimensionless axial velocity decay rate~\cite{pope2000turbulent} where $B=6.0$. The radial mean velocity profile is ${\overline{u}_r(r,z)}/{U_0(z)} = g(\eta)$ and recall $\overline{u}_\theta(r,z) =0$. Using continuity $g(\eta)$ can be expressed in terms of $f(\eta)$. A good approximation for the axial velocity profile is a Gaussian one~\cite{pope2000turbulent}, 

\begin{equation}\label{f_eta}
	f(\eta)= e^{-A_1\eta^2}
\end{equation}
\noindent which is employed for simplicity within this model. This assumption with continuity leads to the radial profile:

\begin{equation}
	\frac{\overline{u}_r(r,z)}{ U_{0}(z)}  = g(\eta) =
	\eta e^{-A_1\eta^2} - \dfrac{1-e^{-A_1\eta^2}}{2A_1\eta}.\label{eq:DefMeanRadialComp} 
\end{equation}

The second-order statistics required to model Lagrangian time-histories in the jet, which have also been found to observe self-similarity~\cite{so1986similarity,panchapakesan1993turbulence,hussein1994velocity,pope2000turbulent}, are represented as:

\begin{align}
	\frac{\sigma_{{z}}^2(r,z)}{U_{0}^2(z)} &= C_2e^{-A_2(\eta-D_2)^2},\label{eq:DefVarAxialComp}\\  \frac{\sigma_{{r}}^2(r,z)}{U_{0}^2(z)} &= C_3e^{-A_3\eta^2},\label{eq:DefVarRadialComp}  \\ \frac{\sigma_{{\theta}}^2(r,z)}{U_{0}^2(z)} &= C_4e^{-A_4(\eta-D_4)^2}\label{eq:DefVarAziComp},
\end{align}

and the cross-correlation is described as,

\begin{equation}
	f_{rz}(\eta) = e^{-A_1\eta^2}\left( \dfrac{1-e^{-A_1\eta^2}}{2A_1\eta}\right)\label{eq:DefFrz}.
\end{equation}

\noindent Note that only the relationship for circumferential velocity variance is included above as the mean is zero. 

Finally, to account for temporal evolution of the jet background properties, recall that we must non-stationarize the time step of the model with the large eddy turnover time scale, $T_E$. This scale has been found to be proportional to the axial position, $z$, within the jet. With the knowledge that the integral length scale, $L$ of a flow field is approximated as $L\approx \sigma_u T_E$ and $\sigma_u \propto 1/z$, we conclude that 

\begin{equation}
	T_{E_z}(z) = C_5\bigg(\frac{D}{U_J}\bigg)\bigg(\frac{z}{D}\bigg)^2,\label{eq:DefBatchTEz}
\end{equation}
\noindent which has been confirmed experimentally \cite{viggiano2021lagrangian,basset2022entrainment}.

\begin{table}[bt!]
	\centering
	\begin{tabular}{cccccccccc}
		$A_1$ & $A_2$ & $A_3$ & $A_4$ & $D_2$ & $D_4$  & $C_2$ & $C_3$ & $C_4$ & $C_5$\\
		\hline
		79.0 & 57.4 & 32.3 & 72.9 & 0.027 & 0.039 & 0.073 & 0.044 & 0.039 & 0.023\\
	\end{tabular}
	\caption{Fits of the experimentally determined self-similarity profiles which are used as inputs into the Batchelor transformation.\label{tab:jetinput}}
\end{table}
The coefficients needed to impose the mean and second-order moments of velocity in the round jet are well-known from many prior experiments and simulations and can be readily obtained from literature. They are shown in Table~\ref{tab:jetinput}.

\section*{Experimental notes}
Experiments were performed in the Lagrangian exploration module (LEM) \cite{zimmermann2010lagrangian} at the \'Ecole Normale Sup\'erieure de Lyon where a vertically-oriented jet of water is injected into the LEM, a convex regular icosahedral tank full of water. The vertical jet, supplied by a pump connected to a reservoir, is ejected upwards into the tank from a round nozzle with a diameter $D = \SI{4}{mm}$ with an exit speed, $U_J=\SI{7}{m}{s^{-1}}$, providing a Reynolds number based on the diameter $Re_D = U_JD/\nu \simeq \SI{2.8e4}{}$ with $\nu$ the water kinematic viscosity. An interrogation volume spanning $\SI{60}{mm} \leq z \leq \SI{180}{mm}$ ($15 \leq z/D \leq 45$) is considered, with $z = 0$ the nozzle exit position. 

The particles are neutrally buoyant spherical polystyrene tracers with a density $\rho_p = \SI{1060}{kg/m^3}$ and a diameter $d_p = \SI{250}{\micro\meter}$. Only the jet is seeded (the quiescent water inside the LEM is not seeded), therefore the particles tracked are from a point source.   Lagrangian particle tracking is performed using three Phantom V12 cameras to create three component, three dimensional trajectories. Backlight illumination is provided by LED panels opposite the cameras. The measurement volume is $80 \times 100 \times \SI{130}{mm^3}$ with a resolution of roughly $\SI{0.1}{mm}$. TTL triggering is employed at a frequence of $\SI{6}{kHz}$ for 8000 snapshots for each run. Two nozzle positions are used to reach from very near the nozzle up to 50 diameters downstream. 50 runs are performed for each nozzle position for convergence of ensemble averaging. For a complete description of the hydraulic and optical set-ups as well as Lagrangian particle tracking and post-processing methods c.f. Ref. \cite{viggiano2021lagrangian}.

\section*{Statistics sampling of instationary signals}
Due to the non-stationarity of the trajectories, conditions need to be applied to accurately characterize and compare statistics of the model output and the experimental results. As depicted in the schematics of Fig.~\ref{fig:TS}, one method to accurately compare the signals is to look at velocity differences of the trajectories conditioned on a initial location within a small sphere with a radius $r_{max}$. The given ensemble is created based on the trajectories that fall within the sphere, namely $X_b$ and $X_c$ from figure~\ref{fig:TS}. The sphere radius, $r_{max}$, is set to the jet half-width divided by two, $\frac{1}{2}R_{1/2}$, to ensure convergence of the statistics while still sampling near the given axial location of interest. Next, statistics are calculated based on the $t_0$ (at the location closed to the center of the interrogation sphere), i.e., the trajectories are now given a pseudo origin at $t_{0b}$ and $t_{0c}$, for $X_b$ and $X_c$, respectively. In an attempt to accurately average over the non-stationary trajectories, each time step is taken from $t_0$.

\begin{figure}[ht!]
	\centering
	\subfigure[]{\includegraphics[width=.5\textwidth]{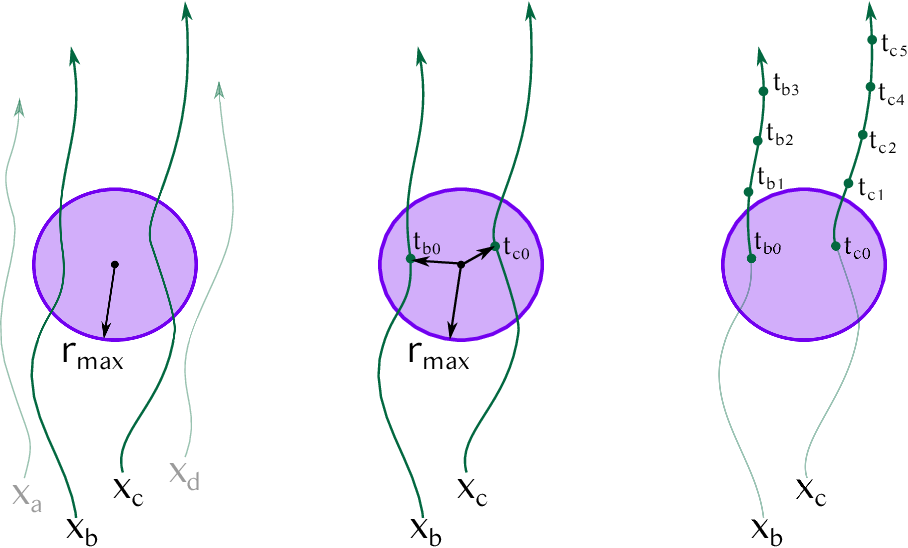}}\hspace{4cm}
	\subfigure[\hspace{-2mm}]{\includegraphics[width=.15\textwidth]{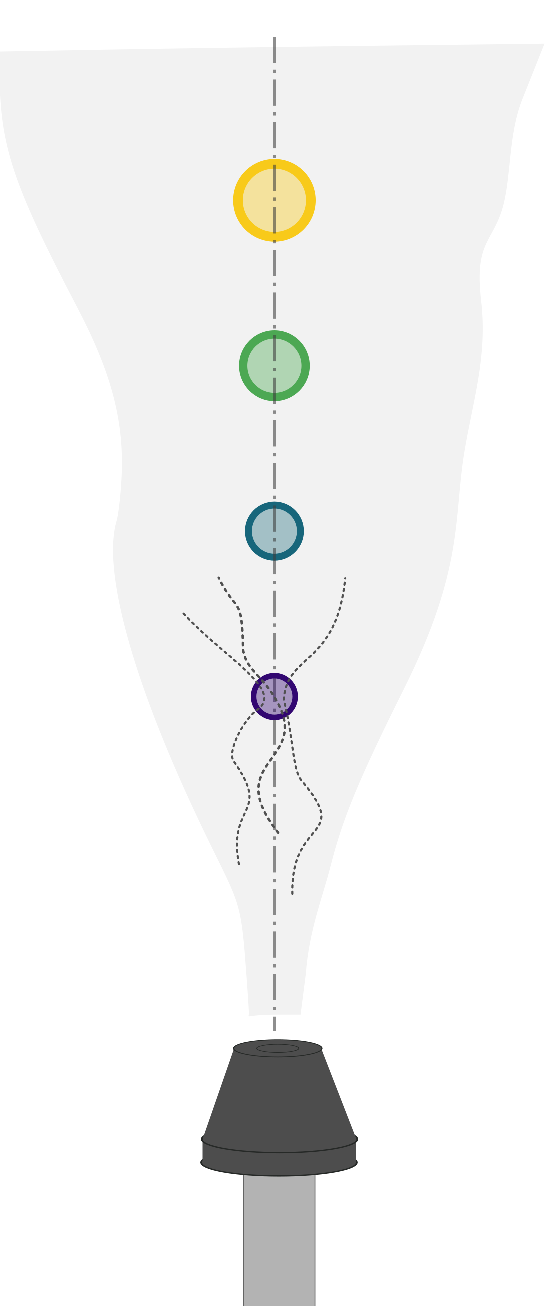}}
	\caption{Schematic of the conditioning for statistical analysis of the non-stationary trajectories.}
	\label{fig:TS}
\end{figure}

\end{widetext}

\end{document}